\documentclass[12pt,a4paper]{article}
\usepackage{t1enc}
\usepackage[latin1]{inputenc}
\usepackage[english]{babel}
\usepackage{graphics}
\begin{document}

\title{AUTOMODULATIONS IN KERR-LENS MODELOCKED SOLID-STATE LASERS}
\author{J. Jasapara$^{ 1 }$, V. L. Kalashnikov$^{ 2 }$, D. O. Krimer$^{ 2}$ , \\
I. G. Poloyko$^{ 2 }$, M. Lenzner$^{ 3 }$, and W. Rudolph$^{ 1 }$}
\maketitle
\begin{center}
$^{1}$Department of Physics and Astronomy, University of New Mexico, Albuquerque,
NM 87131, USA, phone: (505) 277 208, wrudolph@unm.edu,

$^{2}$International Laser Center, 65 Skorina Ave. Bldg. 17, Minsk 220027,
BELARUS, phone: (375-0172)326 286, vkal@ilc.unibel.by, http://www.geocities.com/optomaplev

$^{3}$Department of Quantum Electronics and Laser Technology, Vienna University
of Technology, A-1040 Vienna, Austria

\end{center}
\textbf{Abstract.} Nonstationary pulse regimes associated with self
modulation of a Kerr-lens modelocked Ti:sapphire laser have been studied
experimentally and theoretically. Such laser regimes occur at an intracavity
group delay dispersion that is smaller or larger than what is required for
stable modelocking and exhibit modulation in pulse amplitude and spectra at
frequencies of several hundred kHz. Stabilization of such modulations,
leading to an increase in the pulse peak power by a factor of ten, were
accomplished by weakly modulating the pump laser with the self-modulation
frequency. The main experimental observations can be explained with a round
trip model of the fs laser taking into account gain saturation, Kerr
lensing, and second- and third-order dispersion.

\vspace{1pt}

OCIS: 320.5550, 320.7090, 320.7160, 270.5530

\clearpage

\section{Introduction}

Nonstationary regimes of ultrashort pulse generation in lasers
have gained a great deal of attention in recent years. While the stable
operation of femtosecond lasers is usually the preferred mode of operation,
nonstationary regimes provide unique opportunities to explore the physical
mechanisms leading to the formation and stabilization of ultrashort pulses.
The trend to generate ever shorter pulses from laser oscillators makes it
desirable to characterize the parameter range of stable operation in detail.
This includes the boundaries of such regimes and the processes leading to
nonstationary pulse behavior. On the other hand self- or induced modulation
of such lasers can increase the peak power at a reduced repetition rate and
are thus attractive for pulse amplification or in experiments where
oscillators do not provide sufficient pulse energies.

Various self-starting and induced nonstationary pulse modes in femtosecond
dye and solid-state lasers have been reported, among them cavity dumping
[1], higher-order soliton formation [2, 3], and periodic amplitude
modulations [4]. In recent years, several kinds of periodic pulse amplitude
modulation have been observed in Ti:sapphire lasers after inserting
apertures or adjusting the dispersion [5]. Also cavity dumping [6] and
oscillations between transverse modes [7, 8] has successfully been realized
in Ti:sapphire lasers. In this paper we describe experiments and theoretical
results of automodulations in fs Ti:sapphire lasers that occur with
frequencies typical for relaxation oscillations.

There are two major approaches to describe the complexity of femtosecond
lasers -- the round trip model where the pulse passes through discrete
elements and a continuous model where the action of individual laser
components is distributed uniformly in an infinitely extended hypothetical
material [9]. Both models allow for analytical as well as numerical
approaches. The most successful round trip model was developed more than 20
years ago [10] and relied on the expansion of the transfer functions of the
individual laser elements. It yielded an analytical \textit{sech} solution for the pulse envelope. Later, this model was extended to include self-phase
modulation and dispersion leading to \textit{sech} pulse envelopes and \textit{tanh} phases
[11]. In a stationary regime the pulses reproduce after one or several
roundtrips and thus exhibit features of solitary waves and solitons.

In fs Ti:sapphire lasers, the major pulse shaping mechanisms are
Kerr-lensing, self-phase modulation (SPM) and group delay dispersion (GDD).
A large number of theoretical approaches exist to describe various features
of such lasers [12 -- 16]. The main condition for pulse stability against
the build-up of noise (spontaneous emission) is the negative net-gain
outside the pulse. While in fs dye lasers this is realized by the interplay
of a slow saturable absorber and gain saturation, in KLM solid-state lasers
a fast saturable absorber effect due to Kerr-lensing provides the positive
gain window. Outside the parameter range for stable pulses, periodic and
stochastic pulse modulations have been predicted [17]. Our theoretical
approach in this paper is based on a modified roundtrip model [18] and
tracks the changes of the amplitude and phase parameters of the pulse from
one roundtrip to the next. As we shall see, the inclusion of cavity
third-order dispersion is crucial to describe the nonstationary pulse modes.

In the first part of the paper, we describe two different self-modulation
modes observed at GDD greater and smaller than what is required for stable
modelocking. In the second part we present a theoretical model based on a
stability analysis of the Kerr-lens modelocking that explains the main
experimental findings.

\section{Experimental}

A soft-aperture KLM Ti:sapphire laser was set up in the usual 4-mirror
configuration with a $3\;\,mm$ thick crystal between two focussing mirrors
of $5\;cm$ focal length, two fused silica prisms for dispersion compensation
and a $5\%$ outcoupling mirror. It was pumped with $4.5\;W$, all lines of an
Ar-ion laser. The two distinctly different regimes of self-modulation 
\textit{A} (\textit{B}) were realized by increasing (decreasing) the
insertion of one of the compensating prisms, that is, adding (subtracting)
positive GDD starting from stable modelocking.

To obtain regime \textit{A}, the intracavity prism insertion (glass path) is
increased. The spectrum first broadens and blue-shifts until it exhibits a
distinct edge at around $730\,\;nm$ and a maximum at $800\;\,nm$. A further
increase in the intracavity glass path beyond this point destroys the
modelocking. A perturbation of the cavity (e.g. pushing one of the end
mirrors or rocking one of the prisms) results in a modulation of the pulse
train of the Ti:sapphire laser which can be stabilized by gradually pulling
out the prism somewhat. The resulting pulse train and spectrum are shown in
Figs.\textit{1a} (upper curve) and \textit{1b}. This modulation is similar
to the self Q-switching observed in Ref. [5]. Although the cavity Q is not
switched in a strict sense and the modulations rather resemble relaxation
oscillations we will keep the term self Q-switching. The peak amplitude
under the Q-switch envelope is about ten times higher than the amplitude in
the cw modelocked regime. As can be seen from the figure, the self Q-switch
period is about $5\;\mu s$ and the width of the Q-switch envelope is about $%
1\;\mu s$. The Q-switched pulses are separated by pre-lasing regions. The
average pulse duration without external pulse compression was measured to be
about $130\;fs$ (after $6\;mm$ of SQ1). We compared the second harmonic of
the pulse train with the square of the fundamental signal. From that, the
pulse duration seemed to be constant over the Q-switched pulse and the
prelasing region. Information on the spectral evolution across the
Q-switched pulse was gained by recording the output of a fast photodiode
placed at the exit slit of a monochromator (bandwidth $15\;nm$). The results
shown in Fig. 2 indicate that the spectrum of the pulses under the
Q-switched envelope evolves with time on either side of the central
wavelength.

We observed a second self-modulation mode, \textit{B}, when the cavity prism was
translated so as to decrease the intracavity glass path, that is to increase
the negative GDD of the laser cavity. While translating the prism, the
spectrum moved towards longer wavelengths and began to narrow until, at a
point, the laser jumped into a strongly self-modulated modelocked regime of
operation. The modulated pulse train and pulse spectrum are shown in Fig.%
3\textit{a} and 3\textit{b}. Again, no measurable variation of the pulse
duration over the modulation period and prelasing region was observed. The
pulse duration was about $40\;fs$ (after about $6\;mm$ of SQ1). We also
monitored the modulated pulse train through a monochromator. We found strong
modulations in the spectral intensity at various wavelengths across the
duration of the modulated envelope; some examples are shown in Fig.4. The
spectrum (cf. Fig.3\textit{b}) and the spectral modulations (cf. Fig.4)
across the envelope, bear close resemblance to the observation of
higher-order solitons reported in Refs. [2] and [3].

Phenomenologically, the existence of regime \textit{A} can be explained by relaxation
oscillations and the interplay of SPM and GDD in the laser. Mode \textit{A} arises
when inserting excess glass into the cavity, that is, introducing positive
GDD. According to the soliton model of the laser, a balance of positive
chirp due to SPM and negative chirp due to GDD can result in a steady state
pulse regime. The prism insertion introduces positive GDD which leaves part
of the positive chirp due to SPM uncompensated. SPM is proportional to the
product of nonlinear refractive index and pulse intensity and thus dependent
on the quotient of pulse energy and duration. Hence, with excess glass,
pulses of lower energy but equal pulse duration can still be supported. This
comprises the observed prelasing regime. Because of this low intensity
lasing, the inversion begins to build up in the gain medium until dumped
into the Q-switched pulse. The modulation is driven by relaxation
oscillations. In the leading edge of the Q-switched pulse the intensity of
the modelocked pulses begins to increase while its duration remains
constant. As a result of the higher intensity the SPM begins to generate new
spectral components as indicated by the spectral broadening seen later in
the Q-switched envelope.

Both modulation regimes \textit{A} and \textit{B} were subject to random fluctuations in pulse
length, height, and repetition frequency. To stabilize operational mode \textit{A},
we modulated the pump laser (modulation depth about $5\%$) using an
acousto-optic modulator. The modulator was driven by the amplified output of
a photodiode that monitored the pulse train of the Ti:sapphire laser. Figure
1\textit{a} shows the Q-switched pulse train and the trigger signal for the
modulator. It was also possible to force modulation with an external
oscillator tuned to near the free-running Q-switch frequency of about $%
200\;kHz$. The more than tenfold increase in the pulse energy in the peak of
the Q-switched pulse ( $30\;nJ$) as compared to the ordinary modelocking
makes this mode of operation attractive for subsequent high-repetition rate
pulse amplifiers and for applications that require a somewhat higher energy
than what is available from oscillators.

\section{Theory}

The aim of this section is to discuss some general aspects of pulse regimes
with periodic modulation and then focus on cases that describe our
experimental observations, in particular, regime \textit{A}. To this end we will
proceed in two steps, first, second-order group delay dispersion (GDD) is
taken into account only, and subsequently, we will add a third-order
dispersion term.

Assuming that the change of the pulse envelope during one roundtrip is
small, the laser dynamics can be described by a nonlinear equation of the
Landau-Ginzburg type [10,19]:

\vspace{1pt}

\vspace{1pt}%
\begin{equation}
\frac{a(k,t)}{k}=\left\{ \alpha -l+(1+id)\partial ^{2}/\partial t^{2}+(\sigma -i\beta
)\left| a\right| ^{2}\right\} a(k,t). 
\end{equation}

Here $a(k,t)$ is the electric field amplitude, $k$ is the round trip number, 
$t$ is the local time normalized to the inverse gain bandwidth $t_{g}$ which
for Ti:sapphire is about $2.5fs$, $\alpha$ is the saturated gain coefficient, $l$
is the linear loss, and $d$ is the GDD coefficient normalized to $t_{g}^{2}$
. The factor before the second derivative, $\left( 1+id\right) $ , takes
into account GDD and the bandwidth limiting effect of the cavity. For the
latter we assume that the bandwidth of the gain medium plays the dominant
role. The last summand consists of two parts, a fast absorber (Kerr lensing)
term $\sigma \left| a\right| ^{2}$ and a self-phase modulation term $-i\beta
\left| a\right| ^{2}$ . For both terms we assumed saturating behavior
according to $\sigma =\sigma _{0}(1-\frac{1}{2}\sigma _{0}\left|
a_{0}\right| ^{2})$ and $\beta =\beta _{0}(1-\frac{1}{2}\beta _{0}\left|
a_{0}\right| ^{2})$ , respectively, where $a_{0}$ is the pulse peak
amplitude and $\beta _{0}=\frac{2\pi zn_{2}}{\lambda n}$ $=1.7\times 10^{-12\;}cm^{2}/W$
is the unsaturated SPM coefficient for Ti: sapphire crystall. Throughout the
paper both $|a_{0}|^{2}$ and $|a|^{2}$ are normalized to $5.9\times10^{11}\;W/cm^{2}$%
, then $\beta _{0}$ and $\sigma _{0}$ will be normalized to $1.7\times
10^{-12}\;cm^{2}/W$. The quantity $\sigma _{0}$ plays the role of an inverse
saturation intensity of the fast absorber (Kerr-lensing). The larger $\sigma
_{0}$ the larger is the amplitude modulation compared to the phase
modulation. The quantities $n_{2}$ = $1.3\times 10^{-16}\;cm^{2}/W$ and $%
n=1.76$ are the nonlinear and linear refractive index, respectively, $z=3\;mm
$ is the length of the Kerr medium, and $\lambda =800\;nm$ is the center
wavelength. The magnitude of $\sigma _{0}$ can be controlled by the cavity
alignment and for typical Kerr-lens modelocked lasers $\sigma _{0}$ is $%
10^{-10}\;-\;10^{-12}\;cm^{2}/W$ [20]. The saturation of SPM is due for
example to the next-higher order term in the expansion for the refractive
index, the $n_{4}$ - term. The approximation that the saturation term
depends on the peak intensity rather the instantaneous intensity is
necessary for solving Eq. (1) analytically.

The gain coefficient, during one roundtrip, changes due to depopulation by
the laser pulse, gain relaxation with a characteristic time $T_{31}$ ( $%
3\;\mu s$ for Ti:sapphire), and the pumping process:

\vspace{1pt}

\vspace{1pt}%
\begin{equation}
\frac{d\alpha }{dt}=\sigma _{14}(\alpha _{m}-\alpha )I_{p}/h\nu _{p}-\sigma
_{32}\alpha \left| a\right| ^{2}/h\nu -\alpha /T_{31}. 
\end{equation}

Here $\sigma _{14}$ $=10^{-19}\;cm^{2}$ and $\sigma _{32}$ $=3\times
10^{-19}\;cm^{2}\;$ are the absorption and emission cross-sections of the
active medium, respectively, $\alpha _{m}$ is the unsaturated gain
coefficient, $\nu _{p}$ and $\nu $ are the pump and laser frequencies,
respectively, and $I_{p}$ is the pump intensity. From this equation of the
gain evolution one can derive a relation between the gain after the \textit{%
k+1}-th roundtrip (left-hand side of the equation, primed quantity) and the 
\textit{k}-th roundtrip (right-hand side of equation):

\vspace{1pt}

\vspace{1pt}%
\begin{equation}
\alpha \prime =\alpha \exp (-2\tau \left| a\right| ^{2}t_{p}-\frac{T_{cav}}{%
T_{31}}-U)+\frac{\alpha _{m}U}{(U+T_{cav}/T_{31})}[1-\exp (-\frac{T_{cav}}{%
T_{31}}-U)], 
\end{equation}

where $t_{p}$ is the laser pulse duration normalized to $t_{g}$ and $T_{cav}$
($10\;ns$) is the cavity roundtrip time. The quantity, $\tau ^{-1}$ is the
gain saturation energy fluency $E_{g}$ ($0.82\;J/cm^{2}$) normalized to $%
\frac{\lambda nt_{g}}{2\pi zn_{2}}$ ($1.5\times 10^{-3}\;J/cm^{2}$) for
which we obtain a value of $555$ that was used in our calculations. $U=\frac{%
\sigma _{14}T_{cav}}{h\nu _{p}}I_{p}$ is a dimensionless pump parameter,
which we chose close to $4\times 10^{-4}$ (for a pump laser spot size of $%
100\;\mu m$ this corresponds to about $4\;W$ pump power). When analyzing
stationary regimes where the pulses reproduce after each roundtrip a
steady-state gain coefficient is used that can be obtained by setting $%
\alpha =\alpha \prime $ in Eq. (\textit{3}). In the simulation of
nonstationary regimes the gain coefficient changes from roundtrip to
roundtrip as described by Eq.(\textit{3}). As is known, see for example
[12], Eq. (1) has a quasi-soliton solution of the form

\vspace{1pt}
\begin{equation}
a(t)=a_{0}\exp (i\phi )/\cosh ^{1+i\psi }(t/t_{p}), 
\end{equation}

where $\psi $ is the chirp term, and $\phi $ is the constant phase
accumulated in one cavity roundtrip. To investigate the pulse stability we
used a so-called aberrationless approximation [21], which allows one to
investigate the dependence of the pulse parameters on the roundtrip number $k
$. After substituting the ansatz (4) into Eq. (1) and expanding of obtained
equation in series of $t$ up to third order, we obtain three ordinary
differential equations as coefficients of expansion. The solution of the
obtained system by the forward Euler method relates the unknown pulse
duration $t_{p}$, chirp parameter $\psi $, and peak amplitude $a_{0}$ after
the $k+1$-th roundtrip (left-hand side of the equation, primed quantities)
to the pulse parameters after the $k$-th transit (right hand side of the
equations):

\vspace{1pt}

\vspace{1pt}%
\begin{equation}
t_{p}\prime =t_{p}+\frac{4-7d\psi -3\psi ^{2}+(\phi \psi -2\sigma
a_{0}^{2}-\psi \beta a_{0}^{2})t_{p}^{2}}{2t_{p}^{2}}, 
\end{equation}

\begin{equation}
\psi \prime =(1-2\sigma a_{0}^{2})\psi +(\phi -\beta a_{0}^{2})\psi
^{2}+\phi -3\beta a_{0}^{2}-\frac{5d+3\psi +5d\psi ^{2}+3\psi ^{3}}{t_{p}^{2}%
},  
\end{equation}

\begin{equation}
a_{0}\prime =a_{0}[1+\frac{d\psi -1+(\alpha -l-\sigma a_{0}^{2})t_{p}^{2}}{%
t_{p}^{2}}]. 
\end{equation}

A fourth equation yields the phase delay,

\vspace{1pt}

\vspace{1pt}%
\begin{equation}
\phi =\beta a_{0}^{2}+\frac{d+\psi }{t_{p}^{2}}. 
\end{equation}

Equations (5-8) describe a stationary pulse regime if the
complex pulse envelope reproduces itself after one cavity roundtrip, that
is, the pulse parameters to the left and the right of the equal sign are
identical. Within a certain range of laser parameters such a stationary
regime usually develops after a few thousand roundtrips and the stationary
pulse parameters can be obtained by solving the system of algebraic
equations. In an unstable or periodically modulated pulse regime, the
evolution of the pulses can be followed by calculating the new pulse
parameters after one additional roundtrip and using these results in the
right-hand side of Eqs. (5-8), i.e., as input for the next roundtrip.

We will first study the behavior of the laser when the SPM parameter $\beta
_{0}$ and the self-amplitude parameter $\sigma _{0}$ are varied. Physically
this can be accomplished by changing the focusing into the crystal and by
changing the cavity alignment or by inserting intracavity apertures,
respectively.

Figures 5, 6, and 7 summarize our evaluation of the mode-locking based on
Eqs. (5-8). Figures 5 and 6 show pulse duration and chirp as a function of the
SPM parameter $\beta _{0}$ and the GDD parameter $d$, respectively. The
solid curves describe steady-state regimes where the pulse parameters
reproduce themselves after one roundtrip while the rectangles indicate
various unstable or periodic pulse modes. Note that these areas refer to a
certain part of a curve rather than to a two-dimensional parameter range. In
region \textit{A} regular (periodic) oscillations of the pulse parameters
occur while in region \textit{B} these oscillations are irregular. Region 
\textit{C} denotes the parameter range where there is no solution to Eq. (1)
in the form of ansatz (4). Region \textit{D} describes oscillations that
have a quasi-regular character. Figure 7 depicts the normalized pulse
intensity versus the global time $T=T_{cav}k$ .

Curve 1 of Fig. 5 describes the (hypothetical) situation of zero GDD. For
small SPM, $\beta _{0}$ $<$ $0.016$ (region \textit{A}), the pulse regime is
unstable. A closer inspection, see Fig. 7\textit{a}, reveals regular
oscillations of the pulse amplitude in this parameter range. The Fourier
spectrum of these oscillations shows seven peaks that broaden rapidly if the
pump power is increased. The resulting broad frequency spectrum indicates a
chaotic behavior of the pulse parameters. An example of the strong
dependence of the amplitude modulation on the pump power is illustrated in
Fig. 7, curve \textit{a} and \textit{b}. A further increase of the SPM, $%
0.016<$ $\beta _{0}$ $<$ $1.3$ (region \textit{B}), leads to irregular
oscillations of the pulse parameters, as can be seen in Fig. 7\textit{c}.
Even larger SPM, $\beta _{0}$ $>$ $2.6$, results in stable pulses.

Curve 2 in Fig. 5 is in the presence of GDD. A comparison with curve 1
shows, that GDD stabilizes the pulse generation in the region of small $%
\beta _{0}$, $\beta _{0}$ $<$ $3.9$, however at substantially longer
pulses. We explain the stabilization to be due mainly to a smaller effect of
nonlinearities as a result of the longer pulse duration and the subsequently
lower intensities. In all cases depicted in Figs. 5 and 6, approaching
regions of instability means increasing the pulse energy. This leads to
stronger gain saturation which makes the onset of relaxation-oscillation
driven instabilities more likely.

Figure 6 depicts pulse duration and chirp as a function of the GDD parameter 
$d$ for a constant SPM parameter $\beta _{0}=1$ and different $\sigma _{0}$
, that is, different magnitude of the amplitude modulation term. Stable
pulses over a broad range of the GDD can be expected for larger values of $%
\sigma _{0}$ (curves 1 and 2). There is no stable regime near zero GDD
(solid rectangle). For small values of $\sigma _{0}$ (curve 3) there is a
broad range of unstable pulse behavior (area \textit{D}). Stable pulses
occur at large negative GDD and around zero and positive GDD. Except for a
region near zero GDD, a larger $\sigma _{0}$ results in longer pulses.

In summary, two distinctly different stable pulse modes exist. One where the
effective phase nonlinearity $\beta _{0}$ is large (cf. Fig. 5, curve 1) or
where long pulses and therefore low intensities (cf. Fig. 5, curve 2) result
in strongly chirped output pulses (Fig. 5\textit{b}). In these regimes any
pulse shortening due to the fast absorber action and spectral broadening due
to SPM is counteracted by the bandwidth-limiting element. The second mode
shows the features of solitary pulse shaping, that is the interplay of GDD
and SPM nonlinearity results in a quasi-Schr\"{o}dinger soliton pulse with
small chirp (cf. Fig. 6\textit{b}, curve 3 to the left of region \textit{D}%
). The pulse area agrees with what is predicted from the nonlinear Schr\"{o}%
dinger equation for a fundamental soliton.

At small $\sigma _{0}$ (curve 3, Fig. 6), decreasing the amount of negative
GDD from the stable pulse regime results in periodic oscillations of the
pulse intensity (region \textit{D}). This behavior describes our
experimental observations, cf. Figs. 1 and 2. The calculated temporal
evolution of the pulse amplitude is detailed in Fig. 7\textit{d}. The
Q-switch period is close to the gain relaxation time $T_{31}$, a fact that
supports the notion that this behavior is driven by relaxation oscillations.

The experimental observations were made near zero GDD where higher-order
dispersion effects are more likely to play a role. Therefore, in the next
step, we included a third-order dispersion term in our model, i.e., we added
a term $d_{3}\partial ^{3}/\partial t^{3}$ to Eq. (1). Here $d_{3}$ is a
dimensionless third-order dispersion coefficient, which is normalized $%
t_{g}^{3}$. To solve the resulting differential equation we now make the
ansatz

\vspace{1pt}\vspace{1pt}%
\begin{equation}
a(k,t)=a_{0}sech^{1+i\psi }[(t-\vartheta k)/t_{p}]\/\,\,\/e^{i[\phi k+\omega
(t-\vartheta k)]}, 
\end{equation}

\vspace{1pt}

where $\vartheta $\ is the pulse delay after one full round trip with
respect to the local time frame, and $\omega $ is the frequency detuning
from the center frequency of the bandwidth-limiting element [22]. These two
additional parameters become necessary in order to describe the effect of
third-order dispersion. After inserting the ansatz (9) into the modified
laser equation (1), we now obtain six iterative relations for the pulse
parameters. The first three equations are comprised of Eqs. (5-7) supplemented by the additional terms

\vspace{1pt}

\begin{eqnarray*}
&&-\omega (21d_{3}\psi +\vartheta \psi t_{p}^{2})-d\psi \omega
^{2}t_{p}^{2}-d_{3}\omega ^{3}\psi t_{p}^{2}, \\
&&\omega (3\vartheta +\theta \psi ^{2}-[25+9\psi
^{2}]d_{3}/t_{p}^{2})-\omega ^{2}(6d-5\psi\\
&&-d\psi ^{2}+\phi t_{p}^{2}-\sigma
a_{0}^{2}t_{p}^{2})-\omega ^{3}(10d_{3}-d_{3}\psi ^{2}+\vartheta t_{p}^{2})-
\\
&&d\omega ^{4}t_{p}^{2}-d_{3}\omega ^{5}t_{p}^{2}, \\
&&3d_{3}\omega \psi -\omega ^{2}t_{p}^{2},
\end{eqnarray*}

respectively. To the expression for the phase delay [Eq.(8)] the
term $\omega (3d_{3}\omega /t_{p}^{2}-\vartheta )+d\omega ^{2}+d_{3}\omega
^{3}$ has to be added. The two additional equations for the pulse delay per
roundtrip $\vartheta$ and frequency detuning $\omega $ are

\vspace{1pt}

\vspace{1pt}%
\begin{equation}
\vartheta =\frac{\omega (2-4d\psi +2d^{2}\psi ^{2})+\omega ^{2}d_{3}(9d\psi
^{2}-\psi )+9\omega ^{3}d_{3}^{2}\psi ^{2}-8d_{3}\psi (\alpha -l+\sigma
a_{0}^{2})}{-\psi +d\psi ^{2}+3d_{3}\omega \psi ^{2}}, 
\end{equation}

and

\vspace{1pt}

\vspace{1pt}%
\begin{equation}
\omega \prime =\omega +\frac{8d_{3}\psi -t_{p}^{2}[2\omega +\vartheta \psi
-2d\omega \psi -3d_{3}\psi \omega ^{2}]}{t_{p}^{4}},
\end{equation}

respectively. As is obvious from Eqs. (10) and (11), a nonzero third-order
dispersion term gives rise to a pulse group delay (with respect to the local
time) and a detuning of the carrier frequency from the center of the
bandwidth-limiting element. Figures 8\textit{a} and 8\textit{b} show the
chirp evolution and the frequency shift plotted for the same parameters as
used in Fig. 7\textit{d} and with $d_{3}=-30$. In Fig.7\textit{d} a stable
pulse operation is periodically disturbed by an increase in the pulse
amplitude. Figure 8 is a zoomed-in snapshot describing the chirp and
frequency shift during these bursts. Figure 9 shows in detail the evolution
of the pulse spectrum near the peak of the Q-switched pulse envelope. The
calculated time dependent spectral features are in qualitative agreement
with the experimental observations shown in Fig. 2.

\section{Conclusions}

Kerr-lens modelocked Ti:sapphire lasers can be operated in regimes where
periodic oscillations of the pulse amplitude on a time scale of several
hundred kHz exist. Two such regimes were observed at intracavity group delay
dispersion larger or smaller than what is required for stable modelocking.
Stabilization of the pulse amplitude modulation can be accomplished by
weakly modulating the pump laser with a signal derived from the modulated
Ti:sapphire pulse train. The nature of the self-modulation is similar to
relaxation oscillations well known in solid state lasers. A theoretical
model of the femtosecond laser that takes into account gain saturation,
dispersion, self-phase modulation, and Kerr-lensing explains the trigger of
the self-modulations by the interplay of nonlinear self-phase modulation and
dispersion. In particular, the strong variation of the pulse spectrum
observed during the oscillation spikes can be explained by the theory when
third-order dispersion is included. Depending on the relative contribution
of SPM and GDD two different pulse stabilization mechanisms exist -- the
interaction of strongly chirped pulses with the bandwidth-limiting element,
and the Schr\"{o}dinger-soliton stabilization mechanism for nearly
chirp-free pulses.
The basic calculations in framework of the computer algebra system \textit{Maple} are available on the site http://www.geocities.com/optomaplev.

\section{Acknowledgements}

This work was supported in part by the W. M. Keck Foundation and the
National Science Foundation (PHY-9601890). V.L.K, I.G.P, and D.O.K
acknowledge financial support from the Belorussian Foundation for Basic
Research (F97-256).

\vspace{1pt}

\section{References:}

[1] E. W. Castner, J. J. Korpershoek and D. A. Wiersma , ``Experimental and
theoretical analysis of linear femtosecond dye lasers'', Opt. Comm. 78, 90
(1990).

[2] T.Tsang, `` Observation of high-order solitons from a modelocked
Ti:sapphire laser'', Opt. Lett. 18, 293 (1993).

[3] F. W. Wise, I. A. Walmsley and C. L. Tang, ``Simultaneous formation of
solitons and dispersive waves in a femtosecond dye ring laser'', Opt. Lett.
13, 129 (1988).

[4] V. Petrov, W. Rudolph, U. Stamm and B. Wilhelmi, ``Limits of ultrashort
pulse generation in cw modelocked dye lasers'', Phys. Rev. A 40, 1474
(1989).

[5] Q. R. Xing, W. L. Zhang and K. M. Yoo, ``Self-Q-switched
self-modelocked Ti:sapphire laser'', Opt. Comm. 119, 113 (1995).

[6] A. Baltuska, Z. Wei, M. S. Pshenichnikov, D. A. Wiersma and R. Szip\"{o}%
cs, ``Optical pulse compression to 5 fs at 1 MHz repetition rate'', App.
Phys. B 65, 175 (1997).

[7] D. Cote, H. M. van Driel, ``Period doubling of a femtosecond Ti:
sapphire laser by total mode locking'', Opt. Lett. 23, 715 (1998).

[8] Bolton S. R, Jenks R. A., Elkinton C. N., Sucha G., ``Pulse-resolved
measurements of subharmonic oscillations in a Kerr-lens mode-locked Ti:
sapphire laser'', J. Opt. Soc. Am. B 16, 339 (1999).

[9] J. C. Diels and W. Rudolph, Ultrashort Laser Pulse Phenomena, Academic
Press, San Diego (1996).

[10] H. Haus, ``Theory of mode locking with a fast saturable absorber'',
J. Appl. Phys. 46, 3049 (1975).

[11] D. K\"{u}hlke, W. Rudolph and B. Wilhelmi, IEEE J. Quantum Electron.
19, 526 (1983).

[12] H. A. Haus, J. G. Fujimoto, E. P. Ippen, ``Analytic theory of additive
pulse and Kerr lens mode locking'', IEEE J. Quant. Electr., QE-28, 2086
(1992).

[13] T. Brabec, Ch. Spielmann, P. F. Curley, and F. Krausz, ``Kerr lens
mode locking'', Opt. Lett. 17, 1292 (1992).

[14] J. L. A. Chilla, and O. E. Martinez, ``Spatio-temporal analysis of the
self-mode-locked Ti:sapphire laser'', J. Opt. Soc. Am. B 10, 638 (1993).

[15] V. Magni, G. Cerullo, S de Silvestri and A. Monguzzi, ``Astigmatism in
Gaussian-beam self-focusing and in resonators for Kerr-lens mode locking'', J. Opt. Soc. Am. B 12, 476 (1995).

[16] V. P. Kalosha, M. M\"{u}ller, J. Herrmann, and G. Gatz ``Spatiopemporal model of femtosecond pulse generation in Kerr-lens
mode-locked solid-state lasers'', J. Opt. Soc. Am. B 15, 535 (1998).

[17] V. L. Kalashnikov, I. G. Poloyko, V. P. Mikhailov, D. von der Linde, ``Regular, quasi-periodic and chaotic behavior in cw solid-state Kerr-lens
mode-locked lasers'', J. Opt. Soc. Am. B 14, 2691 (1997).

[18] V. L. Kalashnikov, V. P. Kalosha, I. G. Poloyko, V. P. Mikhailov, M. I.
Demchuk, I. G. Koltchanov, H. J. Eichler, ``Frequency-shift locking of
continuouse-wave solid-state lasers'', J. Opt. Soc. Am. B 12, 2078 (1995).

[19] F. X. K\"{a}rtner, I. D. Jung, and U. Keller, ``Soliton mode locking with
saturable absorbers : theory and experiments'', IEEE J. Selected Topics in
Quant. Electr., 2, 540 (1996).

[20] J. Herrmann, ``Theory of Kerr-lens mode locking: role of self-focusing
and radially varying gain'', J. Opt. Soc. Am. B 11, 498 (1994).

[21] A. M. Sergeev, E. V. Vanin, F. W. Wise, ``Stability of passively
modelocked lasers with fast saturable absorbers'', Opt. Commun., 140, 61
(1997).

[22] V. L. Kalashnikov, V. P. Kalosha, I. G. Poloyko, V. P. Mikhailov, ``New principle of formation of ultrashort pulses in solid-state lasers with
self-phase-modulation and gain saturation'', Quantum Electr., 26, 236
(1996).

\clearpage

\begin{figure}
	\begin{center}
		\includegraphics{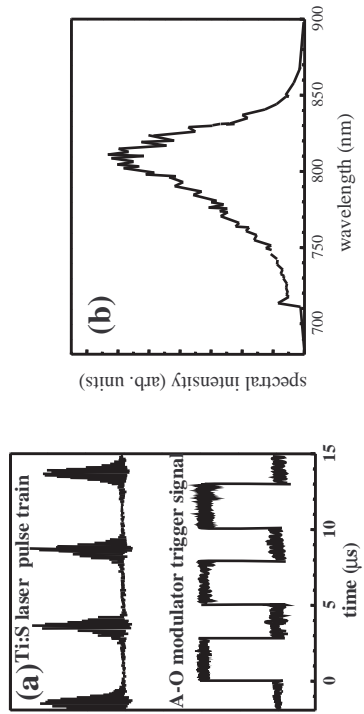}
	\end{center}

	\caption{(\textit{a}) Pulse train from a mode-locked Ti:sapphire laser in a
self Q-switched regime \textit{A} (upper curve) and the acousto-optic driver
signal used in the case when the pump laser was modulated. On the time scale
shown here there was no observable difference between the Ti:sapphire laser
pulse train with the modulated and unmodulated pump laser. (\textit{b})
Laser spectrum observed during regime \textit{A}.}
\end{figure}

\begin{figure}
	\begin{center}
		\includegraphics{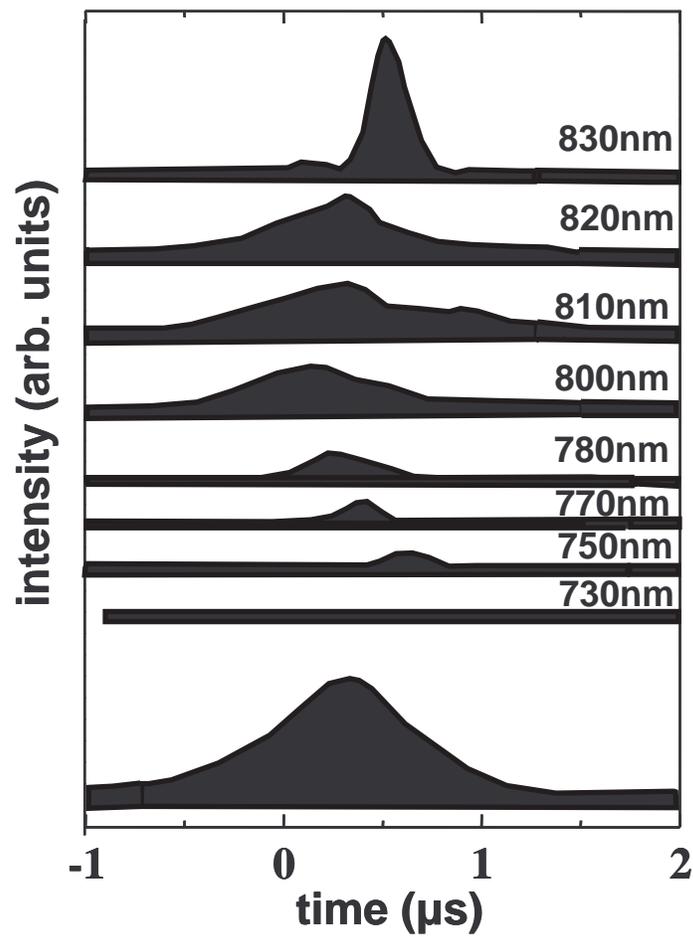}
	\end{center}

	\caption{Temporal position of the various spectral components relative to the peak of the Q-switched envelope and the modelocked pulse train (bottom
curve).}
\end{figure}

\begin{figure}
	\begin{center}
		\includegraphics{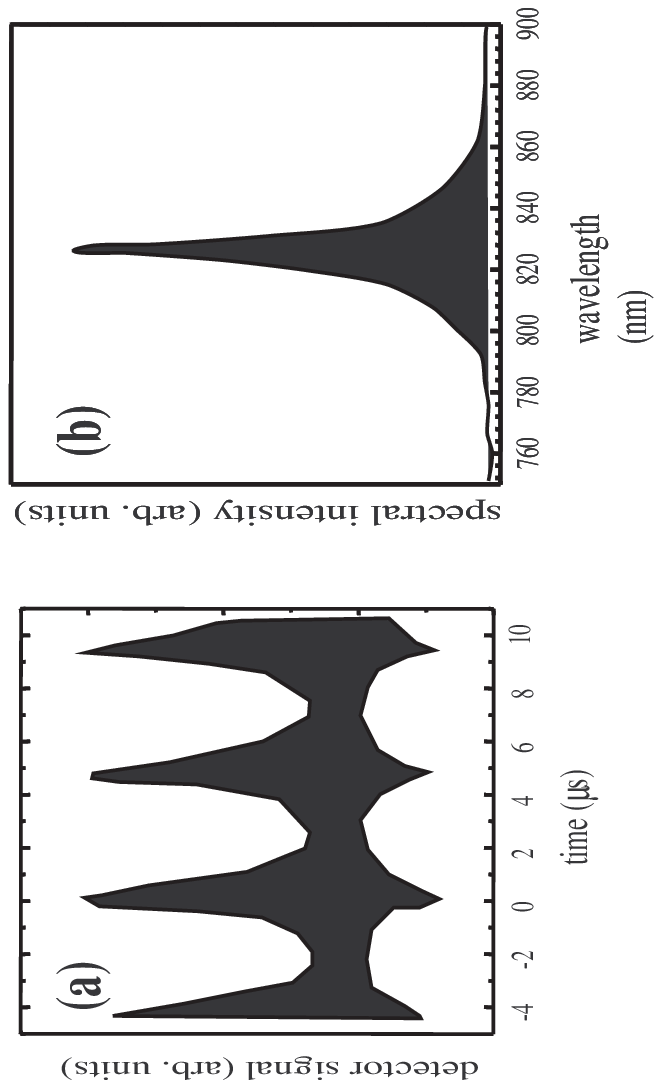}
	\end{center}

	\caption{(\textit{a}) Pulse train and (\textit{b}) spectrum observed in
regime \textit{B}.}
\end{figure}

\begin{figure}
	\begin{center}
		\includegraphics{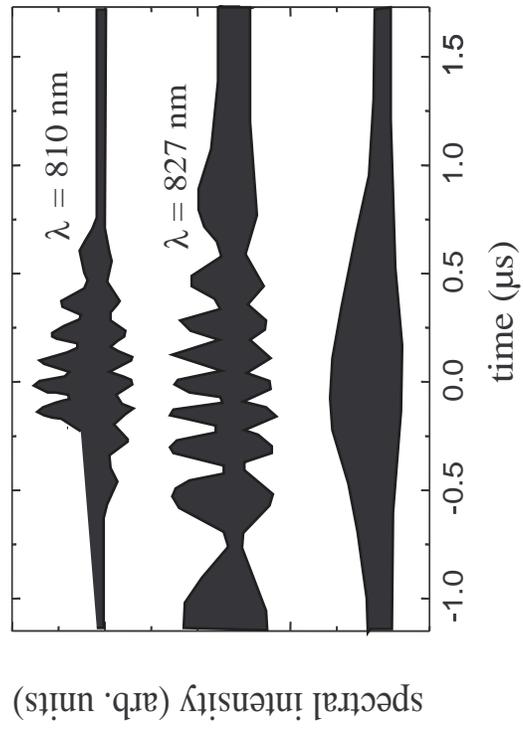}
	\end{center}

	\caption{Temporal behavior of spectral components at $810\;nm$ and $827\;nm$ in regime \textit{B} and the modelocked pulse train (bottom curve).}
\end{figure}

\begin{figure}
	\begin{center}
		\includegraphics{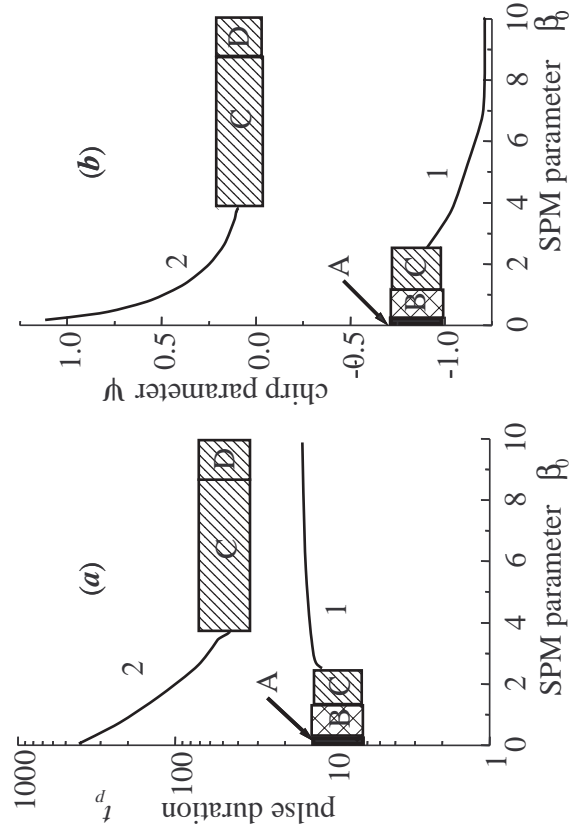}
	\end{center}

	\caption{Steady-state pulse duration (\textit{a}) and chirp (\textit{b})
versus the normalized SPM coefficient $\beta _{0}$ for a normalized GDD
coefficient of $d=0$ (curve 1) and $d=-10$ (curve 2). The other laser
parameters are $\sigma _{0}$ $=1$, $l=0.05$, $\alpha _{m}$ $=0.5$, $%
U=4\times 10^{-4}$, $z=0.3\;cm$, $\lambda =800\;nm$, and $T_{cav}=10\;ns$.
The hatched regions describe various nonstationary pulse regimes, see text.}
\end{figure}

\begin{figure}
	\begin{center}
		\includegraphics{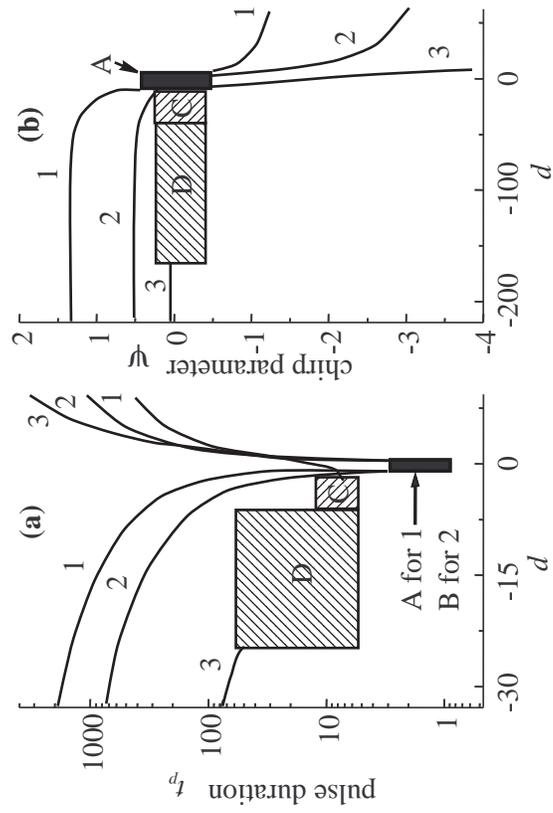}
	\end{center}

	\caption{Steady-state pulse duration (\textit{a}) and chirp (\textit{b})
versus the normalized GDD coefficient $d$ for $\beta _{0}=1$ and $\sigma
_{0}=10$ (curve 1), $\sigma _{0}=1$ (curve 2), $\sigma _{0}=0.1$ (curve 3).
The other parameters are the same as in Fig. 5.}
\end{figure}

\begin{figure}
	\begin{center}
		\includegraphics{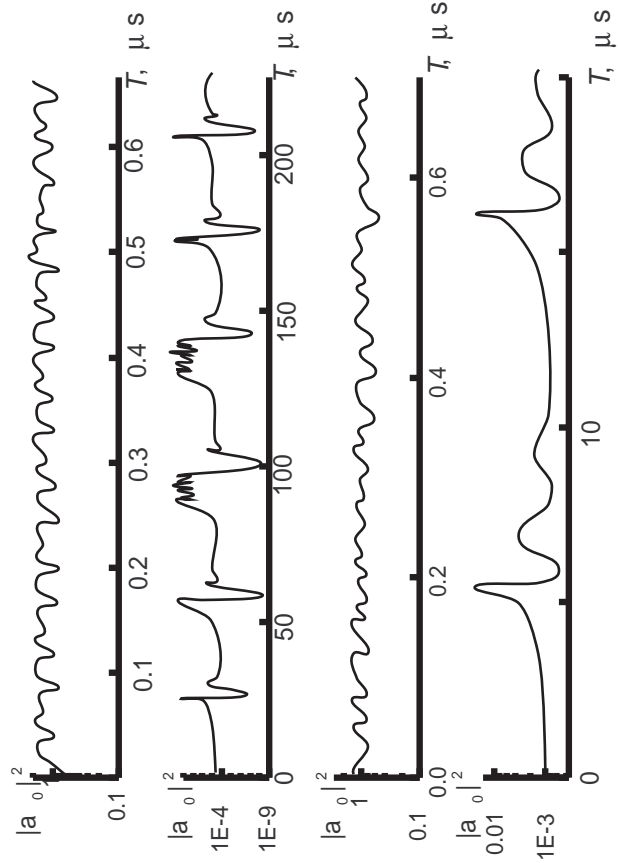}
	\end{center}

	\caption{The normalized pulse intensity $|a_{0}|^{2}$ (log scale) versus
global time $T$. The laser parameters for the different curves are chosen
as follows.
curve \textit{a}: $U=4\times 10^{-4}$, $\beta _{0}=0$, $d=0$, $\sigma _{0}$ $=1$
curve \textit{b}: $U=5\times 10^{-4}$, $\beta _{0}$ $=0$, $d=-10$, $\sigma
_{0}$ $=1$
curve \textit{c}: $U=4\times 10^{-4}$, $\beta _{0}$ $=1$, $d=0$, $\sigma _{0}$ $=1$
curve \textit{d}: $U=6\times 10^{-4}$, $\beta _{0}$ $=1$, $d=-20$, $\sigma
_{0}$ $=0.1$}
\end{figure}

\begin{figure}
	\begin{center}
		\includegraphics{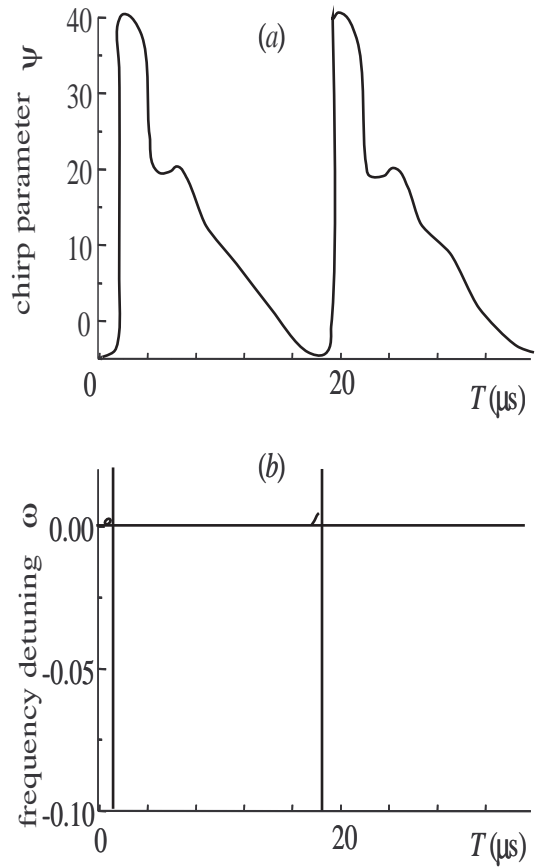}
	\end{center}

	\caption{Pulse chirp $\psi $ (\textit{a}) and normalized frequency shift $%
\omega $ (\textit{b}) versus global time $T$ for the laser parameters used
in Fig. 7, curve \textit{D}, and normalized parameter $d_{3}=-30$.}
\end{figure}

\begin{figure}
	\begin{center}
		\includegraphics{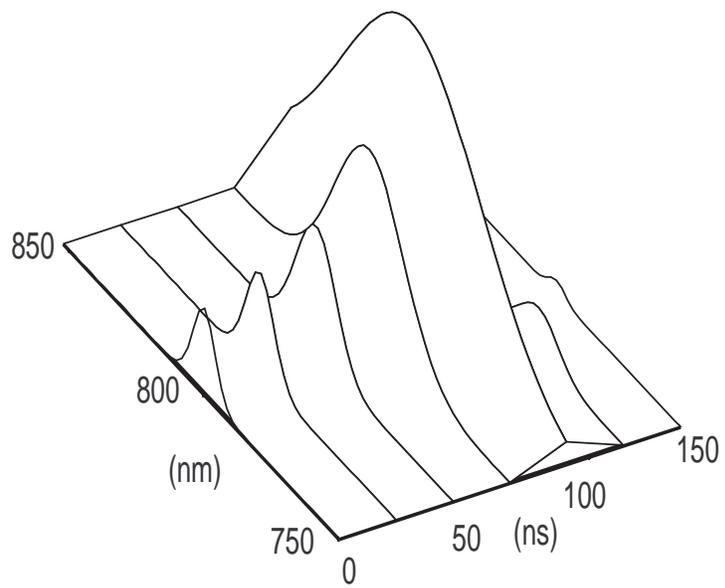}
	\end{center}

	\caption{Calculated pulse spectrum at different times near the peak of the
Q-switched pulse for the parameters of Fig. 8.}
\end{figure}

\end{document}